\documentclass[lettersize,journal]{IEEEtran}
\usepackage{amsmath,amsfonts}
\usepackage{algorithmic}
\usepackage{algorithm}
\usepackage{array}
\usepackage[caption=false]{subfig}
\usepackage{textcomp}
\usepackage{stfloats}
\usepackage{url}
\usepackage{verbatim}
\usepackage{graphicx}
\usepackage[noadjust]{cite}
\usepackage{dsfont}
\usepackage{color}
\usepackage{flushend}
\begin{document}

\title{Differential Communication in Channels with\\ Mobility and Delay Spread using Zak-OTFS}

\author{Sandesh Rao Mattu$^*$, Nishant Mehrotra$^*$, and Robert Calderbank~\IEEEmembership{Fellow,~IEEE}\vspace{-4mm}
        % <-this % stops a space
\thanks{This work is supported by the National Science Foundation under grants 2342690 and 2148212, in part by funds from federal agency and industry partners as specified in the Resilient \& Intelligent NextG Systems (RINGS) program, and in part by the Air Force Office of Scientific Research under grants FA 8750-20-2-0504 and FA 9550-23-1-0249. \\
The authors are with the Department of Electrical and Computer Engineering, Duke University, Durham, NC, 27708, USA (email: \{sandesh.mattu,~nishant.mehrotra,~robert.calderbank\}@duke.edu). \\
$*$ denotes equal contribution.}% <-this % stops a space
%\thanks{Manuscript received April 19, 2021; revised August 16, 2021.}
}



\maketitle

\begin{abstract}
Zak-transform based orthogonal time frequency space (Zak-OTFS) is a delay-Doppler (DD) domain modulation scheme in which the signal processing is carried out in the DD domain. The channel when viewed in the DD domain is predictable. However, even with Zak-OTFS, pilots need to be sent periodically, albeit at a lower rate.
% Channel estimation in wireless communication systems is carried out by transmitting symbols called pilots that are known both at the transmitter and receiver. However these pilot symbols carry no information and also require significant power leading to spectral efficiency loss. The rate at which pilots are transmitted is inversely proportional to how predictable the channel is. 
% Zak-transform based orthogonal time frequency space (Zak-OTFS) is a delay-Doppler (DD) domain modulation scheme in which the signal processing is carried out in the DD domain. The channel when viewed in the DD domain is predictable since the variations in the channel are based on the physics of the reflectors. However, even with Zak-OTFS, pilots need to be sent periodically, albeit at a lower rate. 
In this paper, we propose a differential communication scheme for Zak-OTFS systems that alleviates the need for periodic pilot transmission.
% for data detection and subsequent channel estimation. 
Towards this, we analytically show that the detected data can be used as a pilot and that the channel estimate obtained from the detected data can enable further detection enabling the ``differential'' aspect of the communication. Specifically, we leverage the prediction capability of the DD channel in Zak-OTFS to use the channel estimate (obtained from detected data symbols treated as pilots) in the previous instant to detect data in the next instant and propagate this forward. The advantages are two fold. First, it allows the data symbols to enjoy higher energy since the energy that would otherwise be required for pilot symbols can also be allocated to data symbols. Second, it allows for full spectral efficiency compared to point or embedded pilots. Comparison with the full spectral efficiency achieving spread pilot scheme shows that the proposed method achieves better bit-error rate at lower complexity.
\end{abstract}


\section{Introduction}
\label{sec:intro}

\IEEEPARstart{I}{n} wireless communication systems, information signals are communicated through a wireless medium. These signals undergo various distortions before being received at the receiver \cite{tse2005fundamentals}. To estimate these distortions, the transmitter transmits reference signals called pilots, which are known to the receiver. Typically, these pilot symbols are transmitted at higher energy than the data symbols. The receiver carries out channel estimation, that is, an estimate of the channel is obtained using the received pilot symbols. For a rapidly varying channel, these pilot symbols are transmitted frequently. However, the pilot symbols do not carry any information and therefore frequent pilot transmissions lead to spectral efficiency loss. Different techniques are needed to reduce the frequency of pilot transmissions. Many schemes have been proposed to achieve this~\cite{cai2022joint, tarokh1998new, tarokh2000differential, diggavi2004great}. In this paper we focus on the differential communication scheme.

Differential communication has been studied before in the context of space time block codes (STBC) \cite{tarokh1998new, tarokh2000differential, diggavi2004great}. In \cite{tarokh1998new} the authors use an initial pilot transmission to estimate the channel. Next, assuming that the channel is constant for the next data transmission, the initial channel estimate is used to detect the STBC encoded data. The detected data is modeled as a pilot and used to estimate the channel again which enables more data detection. In \cite{tarokh2000differential}, the authors use differential encoding of information symbols to enable detection at the receiver without the need for a channel estimate. Authors in \cite{diggavi2004great} demonstrate that in addition to the advantages of pilot free communication, differential communication also provides rate advantages at higher layers above the physical layer. In the STBC context, differential communication is possible thanks to the channel becoming predictable in a multi-antenna system \cite{diggavi2004great}. However, this is not readily applicable for single antenna system in a doubly selective channel. New modulation schemes are required for achieving predictability in such communication systems.

Recently, the authors in \cite{bitspaper1, bitspaper2} proposed a delay-Doppler (DD) domain modulation scheme called Zak-transform based orthogonal time frequency space (Zak-OTFS). One of the salient features of Zak-OTFS is that the channel when viewed in DD domain is predictable. This is because the channel changes as fast as the physics of the reflectors would allow, which in practice is slowly varying. With each Zak-OTFS frame typically spanning about few milliseconds, the channel remains almost stationary for a few frame transmission durations.

For channel estimation in Zak-OTFS many DD pilot frames have been proposed \cite{bitspaper2, jayachandran2024zakembedded, Aug2024paper}. These include the point pilot (PP) frame \cite{bitspaper2}, embedded pilot (EP) frame \cite{jayachandran2024zakembedded}, and spread pilot (SP) frame \cite{Aug2024paper}. The  PP frame has a single non-zero value corresponding to the pilot in the whole frame. The EP frame has both pilot and data symbols separated by a guard region to prevent interference between the two. The SP frame has pilot symbols superimposed on the data symbols in way that makes the two mutually unbiased. The PP frame has the least spectral efficiency while SP has full spectral efficiency with EP frame in between. On the other hand, estimation complexity is the highest for SP frame followed by EP and PP frames. 

In this paper, we propose differential communication scheme for doubly-dispersive channel using Zak-OTFS. Our contributions can be summarized as below.
\begin{itemize}
    \item We leverage the predictability of the DD domain channel in Zak-OTFS to design a communication scheme that reduces the frequency of pilot transmissions. 
    \item We propose a differential communication scheme in doubly-dispersive channels using Zak-OTFS. Since all practical channels are doubly-dispersive to various degrees, the method proposed in this paper has practical implications.
    \item We analytically show that the detected data symbols can be used as pilot symbols to estimate the DD channel. We use a time-varying DD channel to show that the estimated channel in a given instant can enable data detection in the subsequent time instant and this process can be repeated over multiple time instants, alleviating the need for frequent pilot transmissions.
    \item The proposed scheme makes for a low-complexity receiver while achieving full spectral efficiency and numerical results demonstrate that the proposed differential communication scheme achieves better bit-error performance when compared to full spectral efficiency achieving SP frame at lower complexity.
\end{itemize}
% In this paper, we propose differential communication scheme for doubly-dispersive channel, i.e., channels with mobility and delay spread, using Zak-OTFS. Since all practical channels are doubly-dispersive to various degrees, the method proposed in this paper has practical implications. We leverage the predictability of the DD domain channel in Zak-OTFS to reduce the frequency of pilot transmissions. We analytically show that the detected data symbols can be used as pilot symbols to estimate the DD channel. We use a time-varying DD channel to show that the estimated channel in a given instant can enable data detection in the subsequent time instant and this process can be repeated over multiple time instants, alleviating the need for frequent pilot transmissions. This makes for a low-complexity receiver while achieving full spectral efficiency. Numerical results demonstrate that the proposed differential communication scheme achieves better bit-error performance when compared to full spectral efficiency achieving SP frame at lower complexity.

\textit{Notation:} $x$ denotes a complex scalar, $\mathbf{x}$ denotes a vector with $n$th entry $\mathbf{x}[n]$, and $\mathbf{X}$ denotes a matrix with $(n,m)$th entry $\mathbf{X}[n,m]$. $(\cdot)^{\ast}$ denotes complex conjugate, $(\cdot)^{\mathsf{T}}$ denotes transpose, and $(\cdot)^{\mathsf{H}}$ denotes complex conjugate transpose. $\langle \cdot, \cdot \rangle$ denotes the inner product between two vectors. 
% Operators acting on complex-valued vectors are denoted by $\mathcal{X}$. 
$\mathbb{Z}$ denotes the set of integers.
% , $\mathbb{Z}_{+}$ the set of positive integers and $\mathbb{Z}_{N}$ the set of integers modulo $N$. 
% $(a,b)$ denotes the greatest common divisor of two integers $a,b$, $\lfloor \cdot \rfloor$ denotes the floor function,
$(\cdot)_N$ denotes the modulo $N$ operation.
% , and $(\cdot)^{-1}_N$ denotes the inverse computed modulo $N$. 
$\delta[\cdot]$ denotes the Kronecker delta function.
$a~ *_\sigma~ b$ denotes the twisted convolution between the DD functions $a$ and $b$. 
% , $\delta(\cdot)$ denotes the Dirac delta function, 
$\mathds{1}_{\{\cdot\}}$ denotes the indicator function, and $\Vert\cdot\Vert_2^2$ denotes the 2-norm of a vector or Frobenius norm of a matrix. $U[a, b)$ denotes a uniform random variable with limits $a$ (inclusive) and $b$ (exclusive), $a<b$.

\section{Preliminaries}
\label{sec:preliminaries}

\subsection{Zak-OTFS}
\label{subsec:zak-otfs}
In Zak-OTFS, each information symbol is mounted on a pulsone which is a pulse train modulated by a tone \cite{bitspaper1, bitspaper2}. A Zak-OTFS frame consists of $MN$ DD bins, where $M$ is the number of delay bins and $N$ is the number of Doppler bins. For a data frame, $MN$ information symbols drawn from a constellation alphabet (e.g., 4-QAM) is mounted on these $MN$ DD bins by modulating each of the $MN$ pulsones. The input-output relation of the Zak-OTFS system \cite{bitspaper2} can be expressed as (we avoid providing a repeat of the system model for brevity)\footnote{In this paper we consider the system model in discrete baseband and quantization and synchronization errors are not considered. However evaluating the performance of the system under these practical non-idealities is an important direction of future research.}:
\begin{align}
    \mathbf{y} = \mathbf{H}\mathbf{x} + \mathbf{n},
\end{align}
where $\mathbf{y} \in \mathbb{C}^{MN\times 1}$ is the vector of received symbols in the DD domain, $\mathbf{H}\in \mathbb{C}^{MN\times MN}$ is the end-to-end channel matrix, $\mathbf{x}\in\mathbb{C}^{MN\times 1}$ is the vector of transmitted symbols in the time-domain, and $\mathbf{n}\in \mathbb{C}^{MN\times 1}$ is the additive Gaussian noise. For detection of $\mathbf{x}$ from $\mathbf{y}$, we need to compute an estimate of $\mathbf{H}$. This is carried out by transmitting pilots. 

\subsection{Channel Estimation}
\label{subsec:channel_estimation}
% There are three main types of pilots used in Zak-OTFS literature. First, is the point pilot (PP) \cite{bitspaper2} which consists of a single non-zero entry in the transmitted DD frame. Second, is the embedded pilot frame, that allows data and pilot to be placed in the OTFS frame, with sufficient guard region around the pilot frame \cite{jayachandran2024zakembedded}, and the spread pilot (SP) frame, where the pilot symbols are superimposed over the data symbols \cite{Aug2024paper} in a way that makes the two unbiased. Regardless of the pilot frame, 
To estimate the channel in a Zak-OTFS system, at the receiver, a DD domain cross-ambiguity function (which also happens to be the maximum-likelihood estimate of the channel \cite{Aug2024paper}) is computed between $\mathbf{y}$ and $\mathbf{x}$. The DD domain cross-ambiguity function is given by:
\begin{align}
    \mathbf{A}_{\mathbf{y}, \mathbf{x}}[k, l] = \sum_{k'=0}^{M-1}\sum_{l'=0}^{N-1}\mathbf{Y}[k', l']\mathbf{X}^*[k'-k, l'-l]e^{-\frac{j2\pi}{MN}l(k'-k)},
    \label{eq:prelim1}
\end{align}
where $\mathbf{Y}[k, l] = \mathbf{y}[k+lM]$ and $\mathbf{X}[k, l] = \mathbf{x}[k+lM]$ are the matrix representations of the corresponding vectors, where $k=0, 1, \cdots, M-1, l = 0, 1, \cdots, N-1$. The estimation of the channel matrix, $\mathbf{H}$ happens in two steps. First, $\mathbf{A}_{\mathbf{y}, \mathbf{x}}$ gives the ``model-free'' \cite{Aug2024paper} estimate, $\widehat{\mathbf{h}}_{\mathrm{eff}}$, of the effective channel matrix $\mathbf{h}_{\mathrm{eff}}$. Next, the estimate $\widehat{\mathbf{h}}_{\mathrm{eff}} \in \mathbb{C}^{M\times N}$ is used to construct the estimated channel matrix $\widehat{\mathbf{H}}$ (see \cite[Eq. (38)]{bitspaper2}). $\widehat{\mathbf{H}}$ is then used for subsequent data detection.

% \subsection{Turbo iterations for SP}
% \label{subsec:turbo_iteration_sp}
% The channel estimate obtained from an SP frame is in the presence of information symbols. To improve the accuracy of the estimated channel and consequently the detected information, turbo iterations are performed \cite{jayachandran2024zakturbo}. This involves moving back and forth between channel estimation and data detection as explained below. Using the initial channel estimation obtained from the spread pilot, the effect of spread pilot is removed. The estimated channel is then used to detect data symbols. The detected data is removed from the received frame to obtain a pilot-only symbol and this is used for estimating the channel again. This back-and-forth estimation and detection is carried for a few iterations for each frame which improves the accuracy of the channel estimation and detected information symbols. 

\subsection{Ambiguity Function of PP}
\label{subsec:amb_fun_pp}
As mentioned earlier, in Zak-OTFS, information symbols are mounted on pulsone bases in the DD domain. The discrete DD domain point pulsone indexed by ($k_0, l_0$) is:
\begin{align}
    \mathbf{X}_{k_0, l_0}^{(\mathrm{p})}[k, l] = \sum_{n, m\in\mathbb{Z}}e^{\frac{j2\pi}{N} nl}\delta[k-k_0-nM]\delta[l-l_0-mN],
    \label{eq:amb_pp_1}
\end{align}
where the term $e^{\frac{j2\pi}{N} nl}$ is to account for quasi-periodicity of the pulsone \cite{Aug2024paper}. Information symbols modulate pulsone at each $(k_0, l_0)$ tuple where $k_0 = 0, 1, \cdots, M-1, l_0=0, 1, \cdots, N-1$. The cross-ambiguity between a pulsone indexed by $(k_0, l_0)$ and $(k_1, l_1)$ is evaluated as (we skip writing the limits for the sum wherever it is clear, for brevity):
\begin{align}
    \mathbf{A}_{\mathbf{X}_{k_0, l_0}^{(\mathrm{p})}, \mathbf{X}_{k_1, l_1}^{(\mathrm{p})}}[k, l] &= \sum_{k_1, l_1}\sum_{n_1, m_1 \in \mathbb{Z}}\sum_{n_2, m_2 \in \mathbb{Z}}e^{\frac{j2\pi}{N}n_1l'} \times \nonumber \\
    &\hspace{5mm}\delta[k'-k_0-n_1M]\delta[l'-l_0-m_1N] \times \nonumber \\
    &\hspace{5mm}e^{-\frac{j2\pi}{N}n_2(l'-l)}\delta[k'-k-k_1-n_2M]\times \nonumber \\
    &\hspace{5mm}\delta[l'-l-l_1-m_2N]e^{-\frac{j2\pi}{MN}l(k'-k)}.
    \label{eq:prelim2}
\end{align}
From $\delta[k'-k_0-n_1M]\delta[l'-l_0-m_1N]$, we have $n_1 = m_1 = 0$ and $k'=k_0, l' = l_0$, since $k_0, k' = 0, 1, \cdots, M-1$ and $l_0, l' = 0, 1, \cdots, N-1$. Substituting in \eqref{eq:prelim2}:
\begin{align}
    \mathbf{A}_{\mathbf{X}_{k_0, l_0}^{(\mathrm{p})}, \mathbf{X}_{k_1, l_1}^{(\mathrm{p})}}[k, l] &= \sum_{k_1, l_1} \sum_{n_2, m_2 \in \mathbb{Z}}e^{-\frac{j2\pi}{N}n_2(l_0-l)} \times \nonumber \\
    &\hspace{5mm}\delta[k_0-k-k_1-n_2M] \times \nonumber \\
    &\hspace{5mm}\delta[l_0-l-l_1-m_2N]e^{-\frac{j2\pi}{MN}l(k_0-k)}.
\end{align}
$\delta[k-(k_0-k_1)-n_2M] \implies k = (k_0 - k_1) \bmod M$ and $\delta[l-(l_0-l_1)-m_2N] \implies l = (l_0 - l_1) \bmod N$. This means that the cross-ambiguity is supported (alternately, the sum is non-zero) on a lattice given by the points $((k_0 - k_1)_M, {(l_0 - l_1)_N})$. The self ambiguity $\mathbf{A}_{\mathbf{X}^{(p)}, \mathbf{X}^{(p)}}[k, l]$ of the pulsone is supported on the lattice $(nM, mN)$ \cite{Aug2024paper} and therefore (up to a phase):
\begin{align}
    \mathbf{A}_{\mathbf{X}_{k_0, l_0}^{(\mathrm{p})}, \mathbf{X}_{k_1, l_1}^{(\mathrm{p})}}[k, l] = \mathbf{A}_{\mathbf{X}^{(\mathrm{p})}, \mathbf{X}^{(\mathrm{p})}}[k - (k_0-k_1), l-(l_0-l_1)].
    \label{eq:prelim3}
\end{align}

% For an information symbol at $(k_0, l_0)$ in the DD grid, the PP basis is $\mathbf{P}_{k_0, l_0}[k, l] = \delta[k-k_0]\delta[l-l_0]$, where $k, k_0 = 0, 1, \cdots, M-1$ and $l, l_0 = 0, 1, \cdots, N-1$. The cross-abmiguity between $\mathbf{P}_{k_0, l_0}[k, l]$ and $\mathbf{P}_{k_1, l_1}[k, l]$ is
% \begin{align}
%     \mathbf{A}_{\mathbf{P}_{k_0, l_0}, \mathbf{P}_{k_1, l_1}}[k, l] = 
% \end{align}
% Using the inverse discrete Zak transform \cite{dzt}, the corresponding time-domain basis is:
% \begin{equation}
%     \mathbf{x}_{(k_0,l_0)}[n] = \frac{1}{\sqrt{N}} \sum_{d \in \mathbb{Z}} e^{\frac{j2\pi}{N} d l_0} \delta[n-k_0-dM],
%     \label{eq:prelim2}
% \end{equation}
% which is a pulse train modulated by a tone. To evaluate the ambiguity function of PP, we proceed as follows. We note that the time-domain ambiguity function is equivalent to the DD domain ambiguity \cite{mehrotra2025zak}. The time-domain ambiguity between two $MN$-length vectors $\mathbf{y}$ and $\mathbf{x}$ is given by:
% \begin{align}
%     \mathbf{A}_{\mathbf{y}, \mathbf{x}}[k', l'] &= \frac{1}{MN}\sum_{n=0}^{MN-1}\mathbf{y}[n]\mathbf{x}^*[n-k']e^{-\frac{j2\pi}{MN}l'(n-k')}.
% \end{align}

\subsection{Channel Realization in DD Domain}
\label{subsec:channel_realization}
To generate the time-varying DD domain channel, we make use of the following equations. For $i$th path ($i=0,1, \cdots, P-1$, for a $P$ path channel) with delay $\tau_i$ and Doppler spread $\nu_i$, the distance between the transmitter and receiver is:
\begin{align}
    d_i(t) = \tau_ic + \frac{\nu_ic}{f_c}t,
    \label{eq:prelim4}
\end{align}
where $f_c$ is the carrier frequency and $c$ is the speed of light. To generate the channel gains we assume a path loss model given by:
\begin{align}
    h_i(t) = \frac{\alpha_i}{d_i(t)} e^{j\theta},
\end{align}
where $\alpha_i$ is a function of power profile of the channel and $\theta \sim \mathrm{U}[-2\pi, 2\pi)$. For a given time $t$, the tuple $(h_i(t), \tau_i, \nu_i)$ characterizes the DD channel.

% \subsection{Differential Detection in STBC}
% \label{subsec:diff_det_stbc}
% The early versions of differential detection using space time block codes (STBC) is proposed in \cite{tarokh1998new}. The idea is to detect the transmitted symbols without estimating the channel repeatedly. To start off differential detection, an initial pilot is transmitted, to estimate the channel. Assuming that the channel is quasi-stationary, the data symbols in the subsequent transmission are detected using this initial channel estimate. Next, the detected symbols are treated as pilots. Using an appropriately chosen STBC allows for a low-complexity and accurate estimate of the channel. This loop is continued and (ideally) no more pilot transmissions are required.

\section{Differential Communication in Zak-OTFS}
\label{sec:diff_det}
In this section, we derive the conditions that enable differential communication in Zak-OTFS.
The time-domain symbols mounted on pulsone bases is:
\begin{align}
    \mathbf{x}[n] = \sum_{k_0, l_0}\mathbf{x}^{(\mathrm{p})}_{k_0, l_0}[n] \mathbf{X}[k_0, l_0],
    \label{eq:dif_det1}
\end{align}
where $\mathbf{x}^{(\mathrm{p})}_{k_0, l_0}[n]$ is the $(k_0, l_0)$th pulsone basis in the time-domain (obtained by the inverse discrete Zak transform \cite{dzt} of \eqref{eq:amb_pp_1}), $\mathbf{X}[k_0, l_0]$ is the $(k_0, l_0)$th data symbol mounted on the corresponding pulsone bases.
The input output relation in time-domain is given by \cite{mehrotra2025spread}:

\begin{align}
    \mathbf{y}[n] = \sum_{k_0, l_0}\sum_{k, l} &\mathbf{h}_{\mathrm{eff}}[k, l]\mathbf{x}^{(\mathrm{p})}_{k_0, l_0}[n-k]e^{\frac{j2\pi}{MN}l(n-k)}\mathbf{X}[k_0, l_0] \nonumber \\
    & + \mathbf{n}[n].
    \label{eq:dif_det2}
\end{align}
The maximum likelihood estimate of the channel is equivalent to computing the cross-ambiguity between the received and transmitted symbols \cite[Appendix H]{Aug2024paper}. Therefore, at the receiver, to estimate the channel, we compute the time-domain cross-ambiguity between the received time-domain symbols (in \eqref{eq:dif_det2}) and the transmitted symbols (in \eqref{eq:dif_det1})%
% \footnote{Note that the time-domain ambiguity function is same as the DD domain ambiguity function, see \cite{mehrotra2025zak} for proof.}
:

\begin{align*}
    \mathbf{A}_{\mathbf{y}, \mathbf{x}}[k', l'] &= \frac{1}{MN}\sum_{n=0}^{MN-1}\mathbf{y}[n]\mathbf{x}^*[n-k']e^{-\frac{j2\pi}{MN}l'(n-k')} \nonumber \\
    &= \frac{1}{MN}\Bigg(\sum_{n=0}^{MN-1}\sum_{k_0, l_0}\sum_{k, l}\mathbf{h}_{\mathrm{eff}}[k, l]\mathbf{x}^{(\mathrm{p})}_{k_0, l_0}[n-k] \times \nonumber \\
    &\hspace{5mm}e^{\frac{j2\pi}{MN}l(n-k)}\mathbf{X}[k_0, l_0]\sum_{k_1, l_1}(\mathbf{x}^{(\mathrm{p})}_{k_1, l_1}[n-k'])^*\times \nonumber \\
    &\hspace{5mm}\mathbf{X}^{*}[k_1, l_1]e^{-\frac{j2\pi}{MN}l'(n-k')} + \nonumber \\
    &\hspace{5mm}\sum_{n=0}^{MN-1}\mathbf{n}[n]\mathbf{x}^*[n-k']e^{-\frac{j2\pi}{MN}l'(n-k')}\Bigg) \nonumber \\
    &\overset{(a)}{=} \frac{1}{MN}\sum_{k, l}\mathbf{h}_{\mathrm{eff}}[k, l]\sum_{k_0, l_0}\mathbf{X}[k_0, l_0] \times \nonumber \\
\end{align*}
\begin{align}
    &\hspace{5mm}\sum_{k_1, l_1}\hspace{-1mm}\mathbf{X}^*[k_1, l_1]\sum_{n=0}^{MN-1}\mathbf{x}^{(\mathrm{p})}_{k_0, l_0}[n-k]e^{\frac{j2\pi}{MN}l(n-k)} \times \nonumber \\
    &\hspace{5mm}(\mathbf{x}^{(\mathrm{p})}_{k_1, l_1}[n-k'])^*e^{-\frac{j2\pi}{MN}l'(n-k')},
    \label{eq:dif_det3}
\end{align}
where step $(a)$ follows because the pulsone samples and noise samples are uncorrelated.
Substituting $\bar{n} = n-k$ in \eqref{eq:dif_det3}:
\begin{align}
    \mathbf{A}_{\mathbf{y}, \mathbf{x}}[k', l'] &= \frac{1}{MN}\sum_{k, l}\mathbf{h}_{\mathrm{eff}}[k, l]\sum_{k_0, l_0}\mathbf{X}[k_0, l_0] \times \nonumber \\
    &\hspace{5mm}\sum_{k_1, l_1}\mathbf{X}^*[k_1, l_1]\sum_{\bar{n}=-k}^{MN-1-k}\mathbf{x}^{(\mathrm{p})}_{k_0, l_0}[\bar{n}]e^{\frac{j2\pi}{MN}l\bar{n}} \times \nonumber \\
    &\hspace{5mm}(\mathbf{x}^{(\mathrm{p})}_{k_1, l_1}[\bar{n}-(k'-k)])^*e^{-\frac{j2\pi}{MN}l'(\bar{n}-(k'-k))} \nonumber \\
    \end{align}
\begin{align}
    &= \sum_{k, l}\mathbf{h}_{\mathrm{eff}}[k, l]\sum_{k_0, l_0}\mathbf{X}[k_0, l_0]\sum_{k_1, l_1}\mathbf{X}^*[k_1, l_1] \times \nonumber \\
    &\hspace{5mm}\mathbf{A}_{\mathbf{x}^{(\mathrm{p})}_{k_0, l_0}, \mathbf{x}^{(\mathrm{p})}_{k_1, l_1}}[k'-k, l'-l]e^{\frac{j2\pi}{MN}l'(k'-k)} \nonumber \\
    &= \mathbf{h}_{\mathrm{eff}}[k', l'] \ \ *_\sigma \nonumber \\
    &\hspace{-5mm}\underbrace{\left(\sum_{k_0, l_0}\mathbf{X}[k_0, l_0]\sum_{k_1, l_1}\mathbf{X}^*[k_1, l_1] \mathbf{A}_{\mathbf{x}^{(\mathrm{p})}_{k_0, l_0}, \mathbf{x}^{(\mathrm{p})}_{k_1, l_1}}[k', l']\right)}_{\mathbf{B}[k', l']}.
    \label{eq:dif_det4}
\end{align}
Since $\mathbf{A}_{\mathbf{x}^{(\mathrm{p})}_{k_0, l_0}, \mathbf{x}^{(\mathrm{p})}_{k_1, l_1}}[k', l'] = \mathbf{A}_{\mathbf{X}_{k_0, l_0}^{(\mathrm{p})}, \mathbf{X}_{k_1, l_1}^{(\mathrm{p})}}[k, l]$ (i.e., the time-domain ambiguity is same as the DD domain ambiguity), substituting \eqref{eq:prelim3} in $\mathbf{B}[k', l']$, we have:
% For a point pulsone, the cross-ambiguity function is $\mathbf{A}_{\mathbf{x}^{(\mathrm{p})}_{k_0, l_0}, \mathbf{x}^{(\mathrm{p})}_{k_1, l_1}}[k, l] = \mathbf{A}_{\mathbf{x}^{(\mathrm{p})},\mathbf{x}^{(\mathrm{p})}}[k-(k_0-k_1), l-(l_0-l_1)]$, where $\mathbf{A}_{\mathbf{x}^{(\mathrm{p})},\mathbf{x}^{(\mathrm{p})}}[k, l]$ is the self-ambiguity of a point pulsone. Therefore, the term in the parenthesis in \eqref{eq:dif_det4} is:
\begin{align}
    \mathbf{B}[k, l] &= \sum_{k_0, l_0}\mathbf{X}[k_0, l_0]\sum_{k_1, l_1}\mathbf{X}^*[k_1, l_1] \times \nonumber \\
    &\hspace{5mm}\mathbf{A}_{\mathbf{X}^{(\mathrm{p})},\mathbf{X}^{(\mathrm{p})}}[k-(k_0-k_1), l-(l_0-l_1)].
    % &= \sum_{k, l}\mathbf{h}_{\mathrm{eff}}[k, l]\sum_{k_1, l_1} \mathbf{X}[k_1, l_1]\mathbf{X}^*[(k_1-(k'-k))_M, (l_1-(l'-l))_N]\mathbf{A}_{\mathbf{x}^{(\mathrm{p})},\mathbf{x}^{(\mathrm{p})}}[k'-k, l'-l]e^{\frac{j2\pi}{MN}l(k'-k)}.
    \label{eq:dif_det5}
\end{align}
Note $\mathbf{A}_{\mathbf{X}^{(\mathrm{p})},\mathbf{X}^{(\mathrm{p})}}[k-(k_0-k_1), l-(l_0-l_1)] = \mathds{1}_{\{k=(k_0-k_1)\bmod M, l=(l_0-l_1)\bmod N\}}$. Substituting $k_1 = (k-k_0)_M$ and $l_1 = (l-l_0)_N$ in \eqref{eq:dif_det5}:
\begin{align}
    \mathbf{B}[k, l] &= \sum_{k_0, l_0}\mathbf{X}[k_0, l_0]\mathbf{X}^*[(k-k_0)_M, (l-l_0)_N].
    \label{eq:dif_det6}
\end{align}
This implies that the cross-ambiguity between the transmitted data frame and the received frame is the twisted convolution between the effective channel $\mathbf{h}_{\mathrm{eff}}[k, l]$ and the inner product between the transmitted data and its delay and Doppler shifted version. If the information symbols are chosen uniformly at random from a constellation, asymptotically we have:
\begin{align}
    \mathbf{A}_{\mathbf{y}, \mathbf{x}}[k', l'] \approx \mathbf{h}_{\mathrm{eff}}[k', l'] *_\sigma e_{\mathrm{d}}\delta[k]\delta[l] = e_{\mathrm{d}}\mathbf{h}_{\mathrm{eff}}[k', l'],
    \label{eq:dif_det7}
\end{align}
where $e_\mathrm{d}$ is the energy of information symbols. The cross-ambiguity between the received and transmitted data symbols is therefore approximately equal to the estimate of the effective channel up to a scale. \\
\textit{Remark 1:} Notice that in the case of spread pilot (\cite{Aug2024paper} or \cite{preamblepaper}), $\mathbf{B}[k, l]$ is exactly $\delta[k]\delta[l]$ and therefore the cross-ambiguity $\mathbf{A}_{\mathbf{y}, \mathbf{x}}[k, l]$ provides the exact estimate of $\mathbf{h}_{\mathrm{eff}}[k, l]$.

% For a DD domain CAZAC array $\mathbf{X}[k, l], \mathbf{B}[k, l] = \delta[k, l]$, and hence $\mathbf{A}_{\mathbf{y}, \mathbf{x}}[k, l] = \mathbf{h}_{\mathrm{eff}}[k, l]$ (which is the case for spread pilot).

\section{Numerical Results}
\label{sec:numerical_results}

\begin{figure*}
    \subfloat[{Actual $\mathbf{h}_{\mathrm{eff}}$}]{\includegraphics[width=0.32\linewidth]{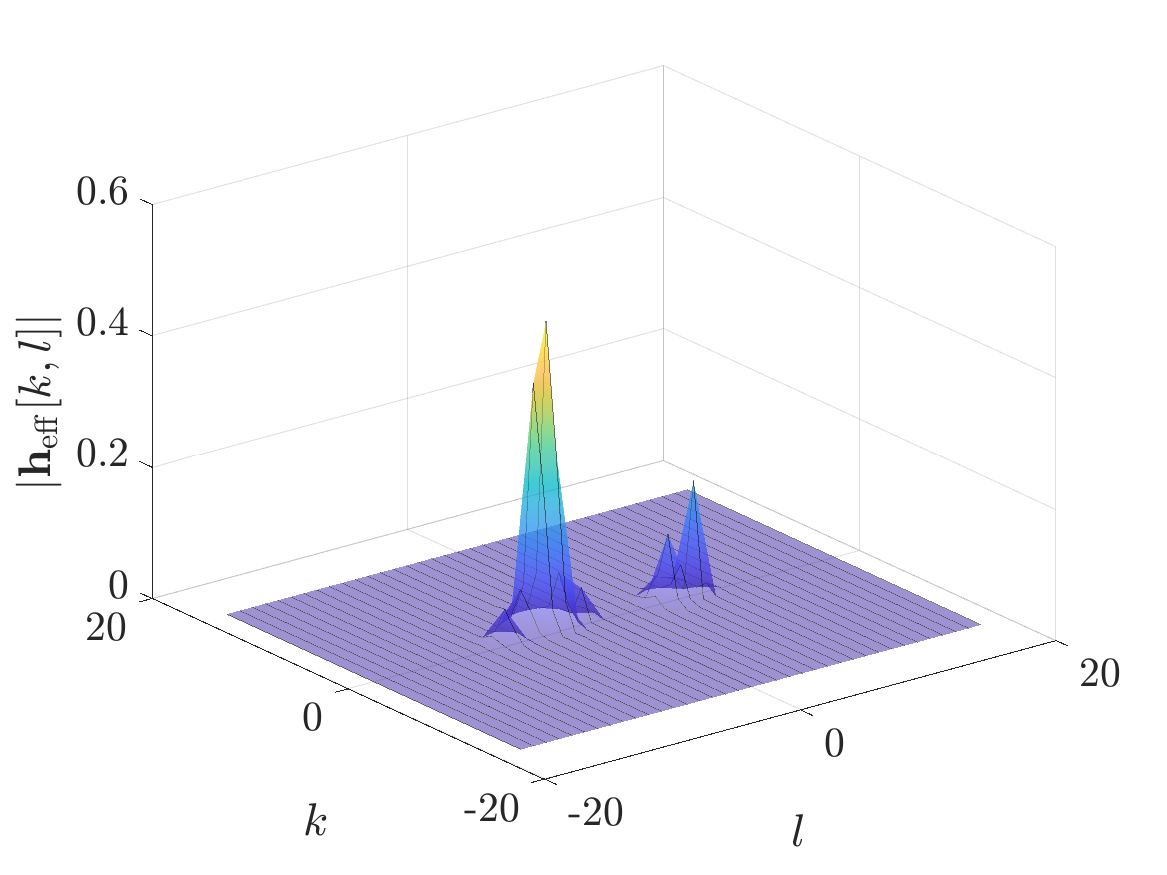}\label{fig:perfect_csi}}
    \hfill
    \subfloat[{$\mathbf{h}_{\mathrm{eff}}$ estimated using SP \cite{preamblepaper}}]{\includegraphics[width=0.32\linewidth]{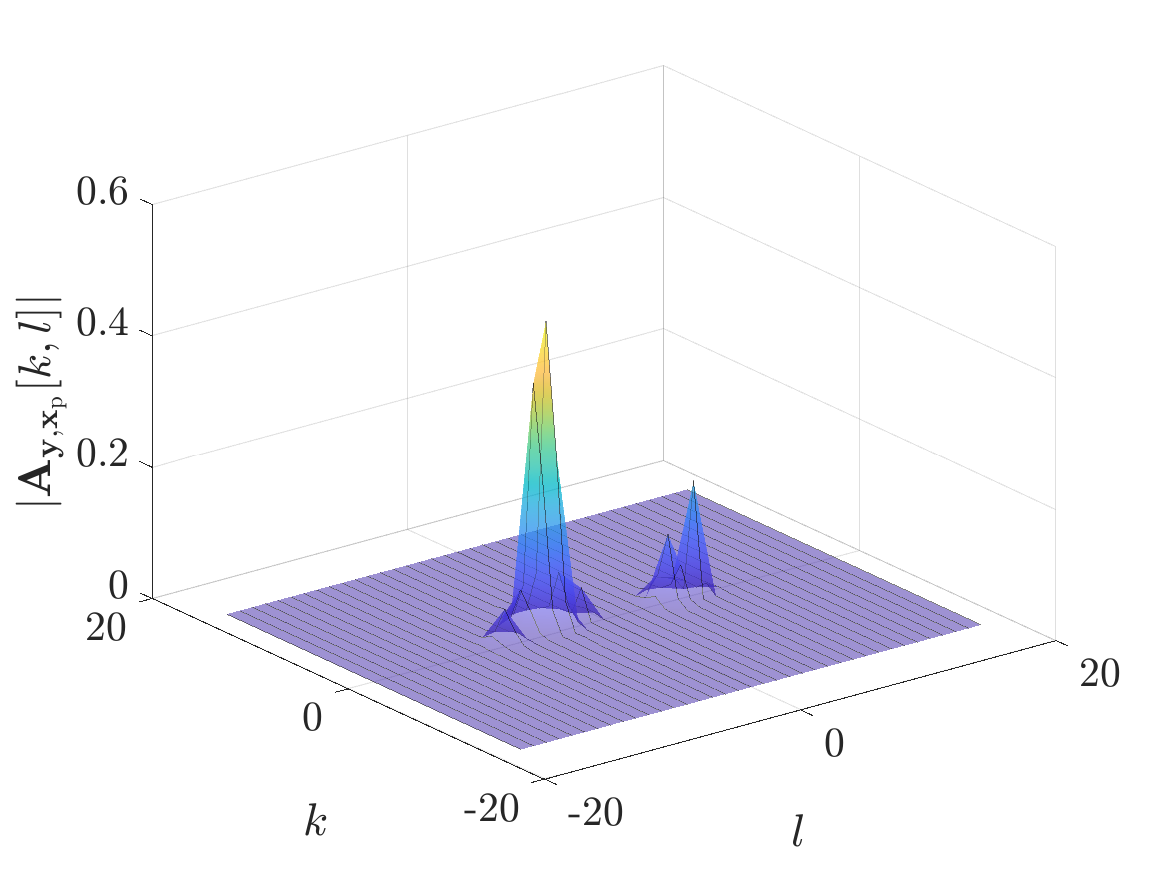}\label{fig:sp_est_csi}}
    \hfill
    \subfloat[{$\mathbf{h}_{\mathrm{eff}}$ estimated using data}]{\includegraphics[width=0.32\linewidth]{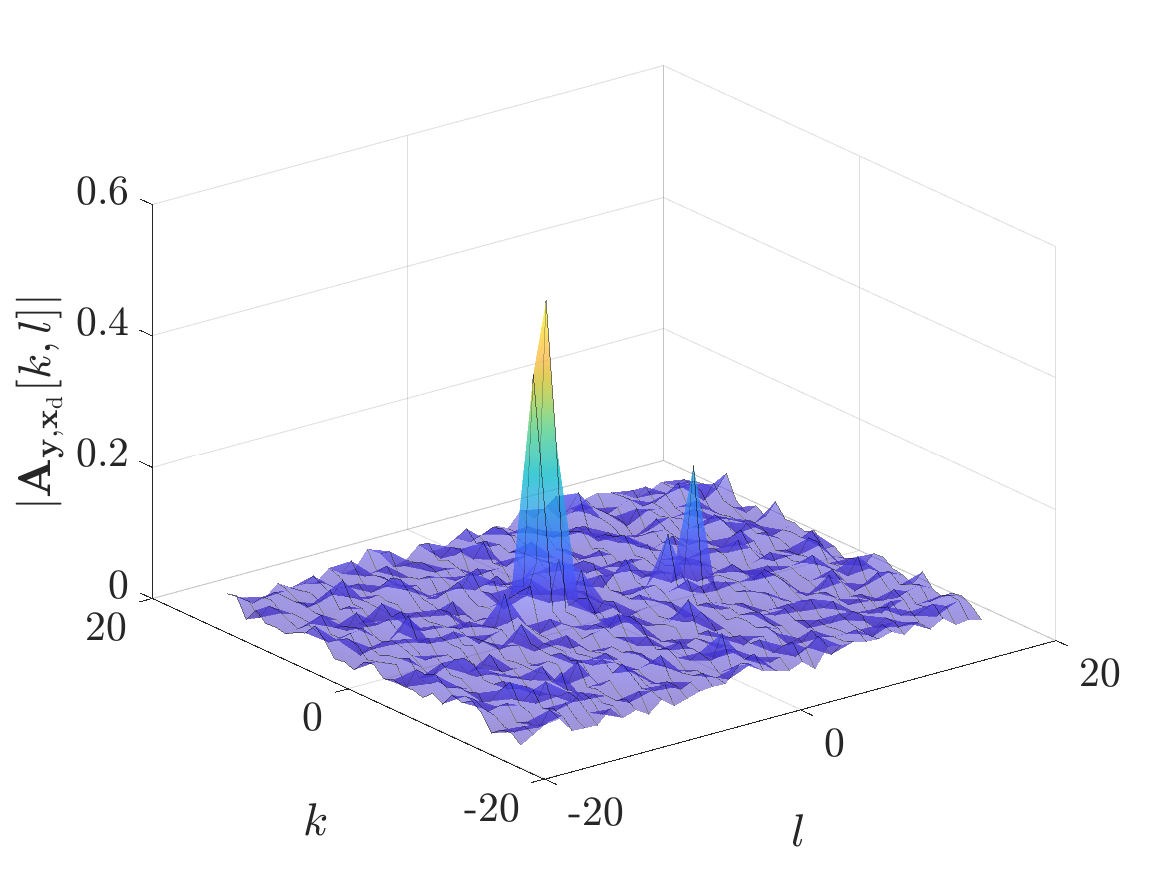}\label{fig:data_est_csi}}
    \caption{Demonstrating differential communication scheme using data symbols as pilots. Perfect channel and channel estimated using pilots are added for reference. Noiseless system. Zak-OTFS system with $M=31, N=37, \nu_p=30 $ kHz, root raised cosine (RRC) pulse shaping with parameter $\beta_\tau=\beta_\nu=0.6$. For the DO frame, data symbols are drawn from a 4-QAM constellation.}
    % \vspace{-4mm}
    \label{fig:estimation_illustration}
\end{figure*}
\begin{table*}
    \centering
    \caption{Table showing a comparison between various schemes in literature.}
    \label{tab:comp_schemes}
    \begin{tabular}{|c|c|c|c|}
        \hline
        Method & Complexity & Operation at receiver & Spectral efficiency\\
        \hline
        \textbf{Differential communication (Ours)} & $\mathbf{\mathcal{O}(M^2N^2)}$ & \textbf{Cross-ambiguity} & \textbf{Full}\\
        \hline
        Spread pilot~\cite{Aug2024paper}\cite{preamblepaper} & $\mathcal{O}(M^2N^2)$ & Cross-ambiguity & Full\\
         & $\mathcal{O}(M^2N^2)$ & Pilot removal & \\
        \hline
        Separate sensing and communication~\cite{bitspaper2} & $\mathcal{O}(M^2N^2)$ & Cross-ambiguity & Half\\
        \hline
    \end{tabular}
\end{table*}
\begin{figure}
    \centering
    \includegraphics[width=\linewidth]{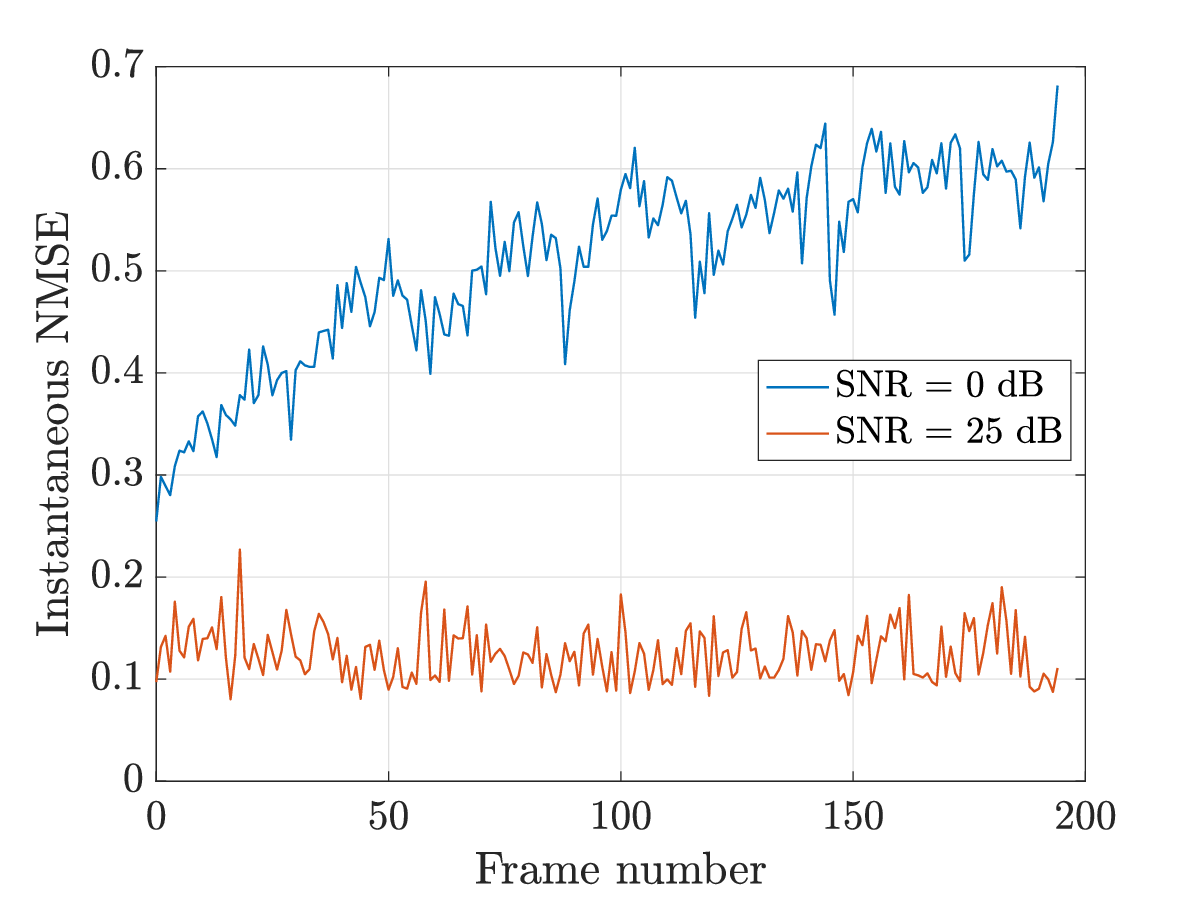}
    \caption{Instantaneous NMSE as a function of number of frames. DO frame with information symbols drawn from 4-QAM. For SNR = 0 dB, the instantaneous NMSE reduces every time a pilot frame is transmitted. For SNR = 25 dB, the number of errors is small and there is no significant error propagation.}
    \label{fig:inst_nmse_no_of_frames}
\end{figure}

We compare the complexity of the proposed differential communication approach against other approaches for Zak-OTFS in literature. The comparison is presented in Table \ref{tab:comp_schemes}. The complexity associated with the proposed approach is $\mathcal{O}(M^2N^2)$ which is incurred from the cross-ambiguity computation, while that for the spread pilot is twice this because of the extra step of pilot removal from the received frame. Using separate frame for sensing and communication incurs the same complexity as the proposed approach but has poor spectral efficiency since one frame is dedicated to the pilot transmission.

Figure \ref{fig:estimation_illustration} compares the estimate of effective channel matrix obtained from SP (Fig. \ref{fig:sp_est_csi}) and from data (Fig. \ref{fig:data_est_csi}) using the proposed approach against the actual channel (Fig. \ref{fig:perfect_csi}). A noiseless system is considered for demonstration purposes. The estimate obtained using data has undulations while the estimate from SP is free from the non-ideality. This corroborates the derivation from \eqref{eq:dif_det6}, where for a data frame, $\mathbf{B}[k, l]$ is not a scaled delta function but for a SP frame it is (see \textit{Remark 1}).

We provide the bit-error rate (BER) and normalized mean square error (NMSE) performance for the proposed differential communication scheme. For the simulations, we consider the Vehicular-A (VehA) channel \cite{veh_a} and generate channel realizations per Sec. \ref{subsec:channel_realization}. The NMSE is computed as:
\begin{align}
    \text{NMSE} = \frac{\Vert\widehat{\mathbf{h}}_{\mathrm{eff}} - {\mathbf{h}}_{\mathrm{eff}}\Vert_2^2}{\Vert{\mathbf{h}}_{\mathrm{eff}}\Vert_2^2},
\end{align}
where $\widehat{\mathbf{h}}_{\mathrm{eff}}$ is the estimated effective channel matrix.

\textit{Note on SNR computations:} We consider an SP frame with pilot energy $e_\mathrm{p, dB} = e_\mathrm{d, dB} - 5 $ dB where $e_\mathrm{d, dB}$ is the data energy in dB scale. The noise variance $\sigma^2 = 1$ and in the linear scale, $e_\mathrm{d, lin} = 10^{\mathrm{SNR_{d}}/10}$, $\mathrm{SNR_{d}}$ is the data SNR. For the data-only (DO) frame, we consider $e_\mathrm{do, lin} = e_\mathrm{p, lin} + e_\mathrm{d, lin}$ to be the energy in the data symbols. Both the SP and DO frames, therefore, carry equal energy, for fair comparison. For all the DO simulations presented below, we transmit a pilot frame every 30 frame transmissions to curb error propagation. 

\begin{figure*}
    \subfloat[{Actual $\mathbf{h}_{\mathrm{eff}}, M=11, N=13$}]{\includegraphics[width=0.32\linewidth]{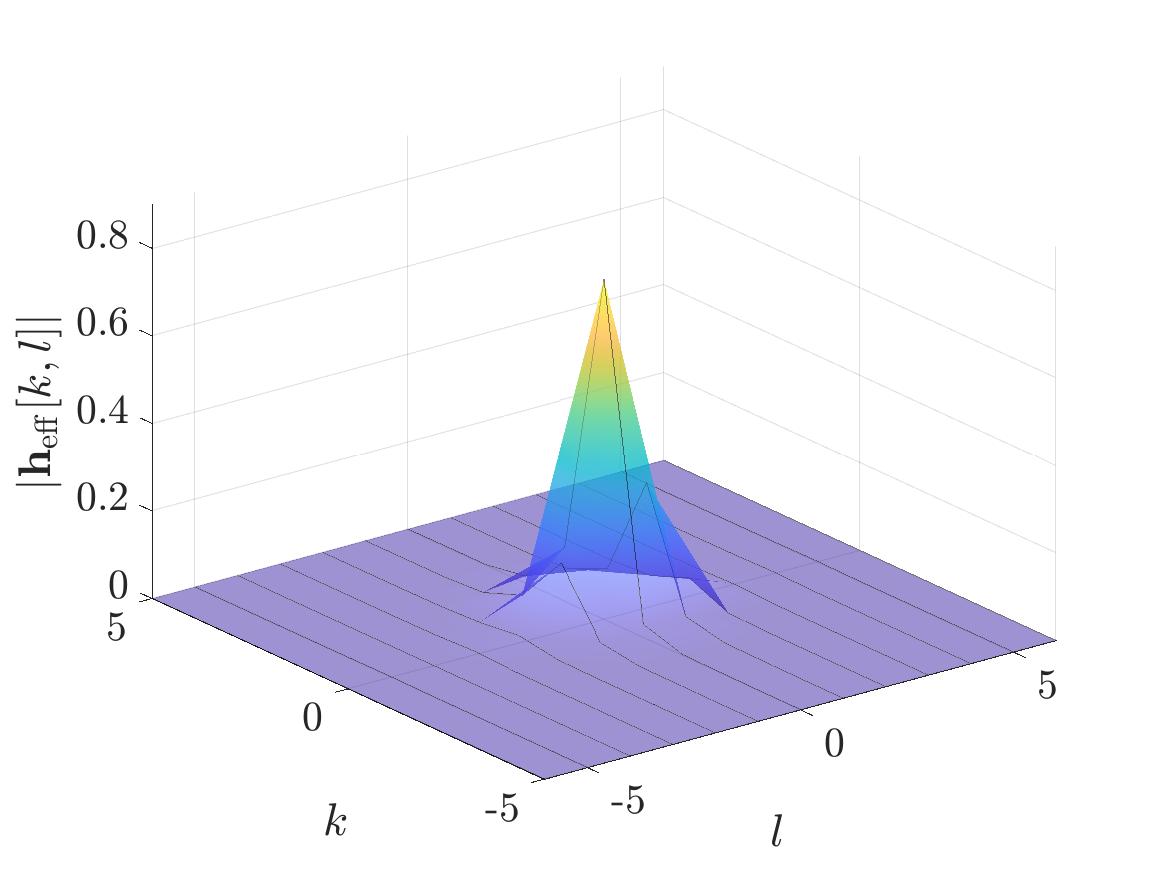}\label{fig:perfect_csi_11_13}}
    \hfill
    \subfloat[{Estimated $\mathbf{h}_{\mathrm{eff}}$ using DO with 4-QAM}]{\includegraphics[width=0.32\linewidth]{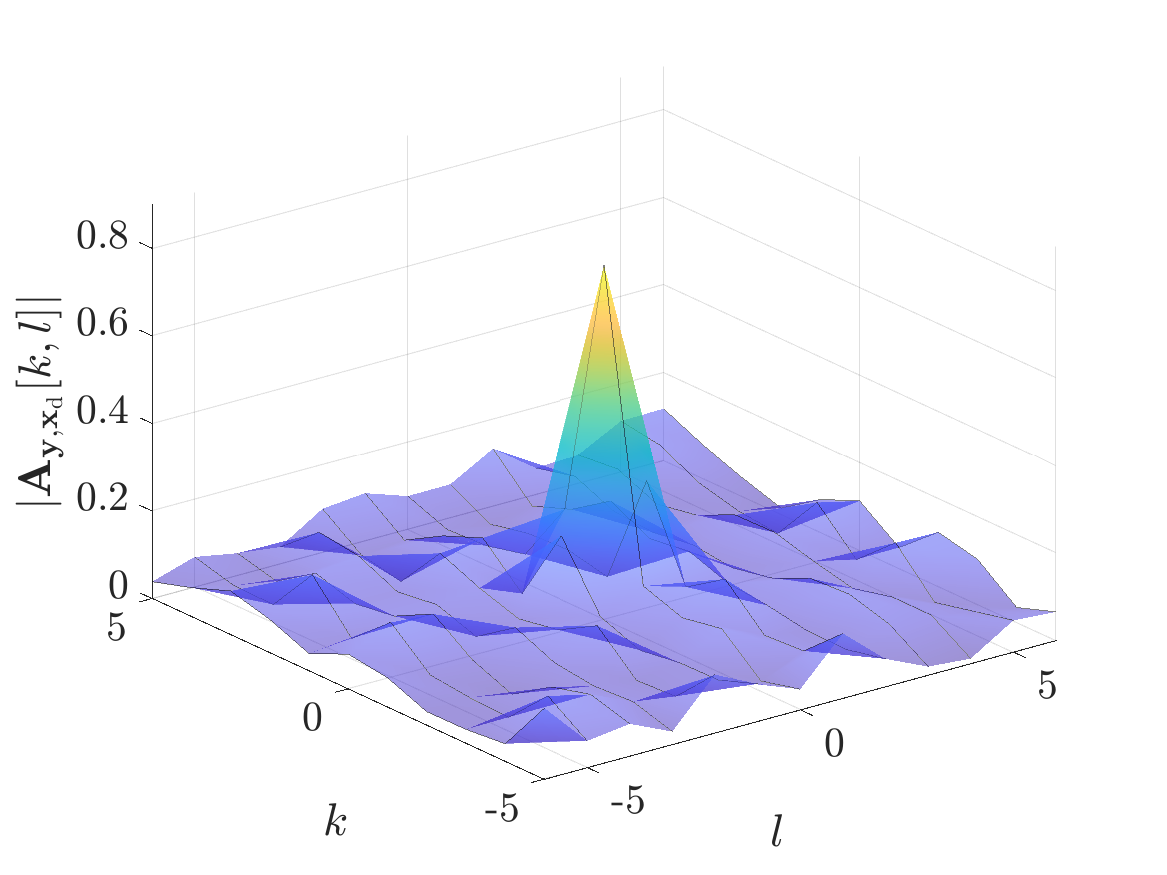}\label{fig:est_csi_11_13_4_qam}}
    \hfill
    \subfloat[{Estimated $\mathbf{h}_{\mathrm{eff}}$ using DO with 256-QAM}]{\includegraphics[width=0.32\linewidth]{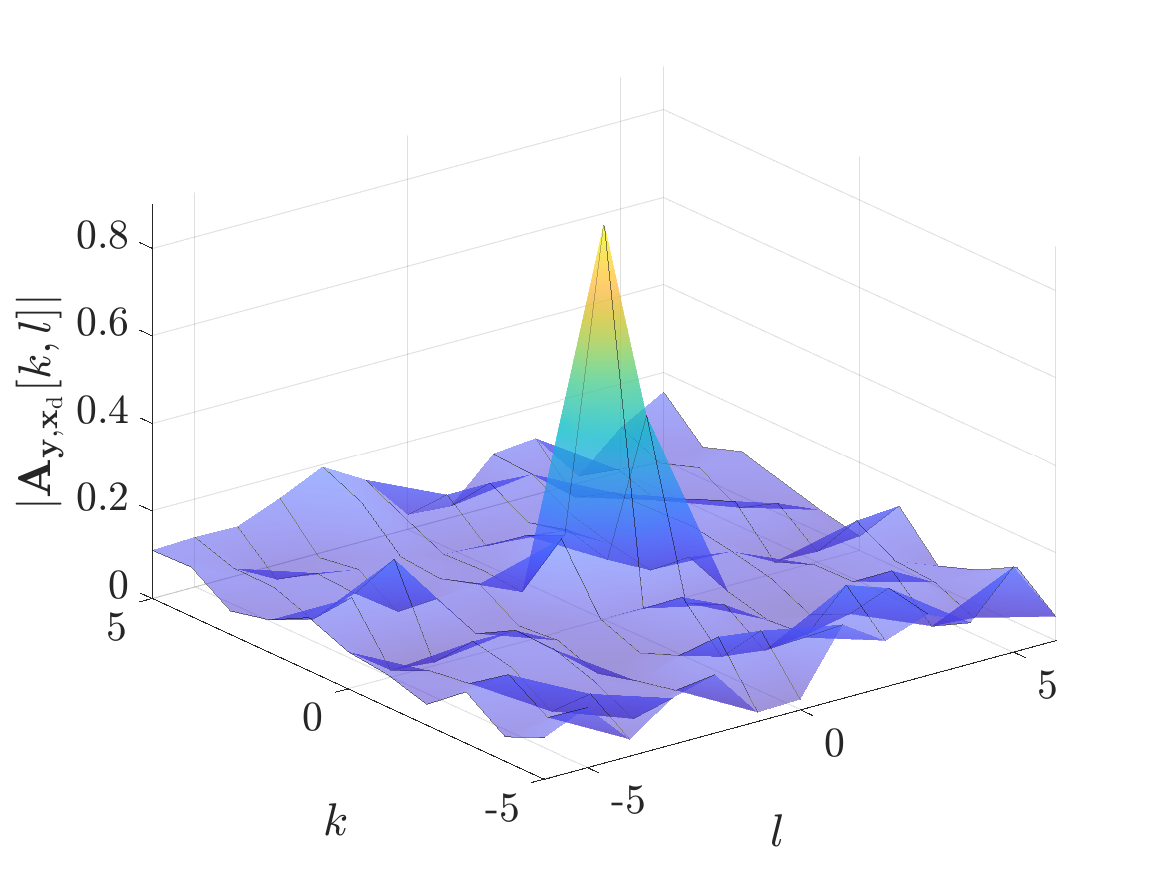}\label{fig:est_csi_11_13_256_qam}}\\
    \subfloat[{Actual $\mathbf{h}_{\mathrm{eff}}, M=31, N=37$}]{\includegraphics[width=0.32\linewidth]{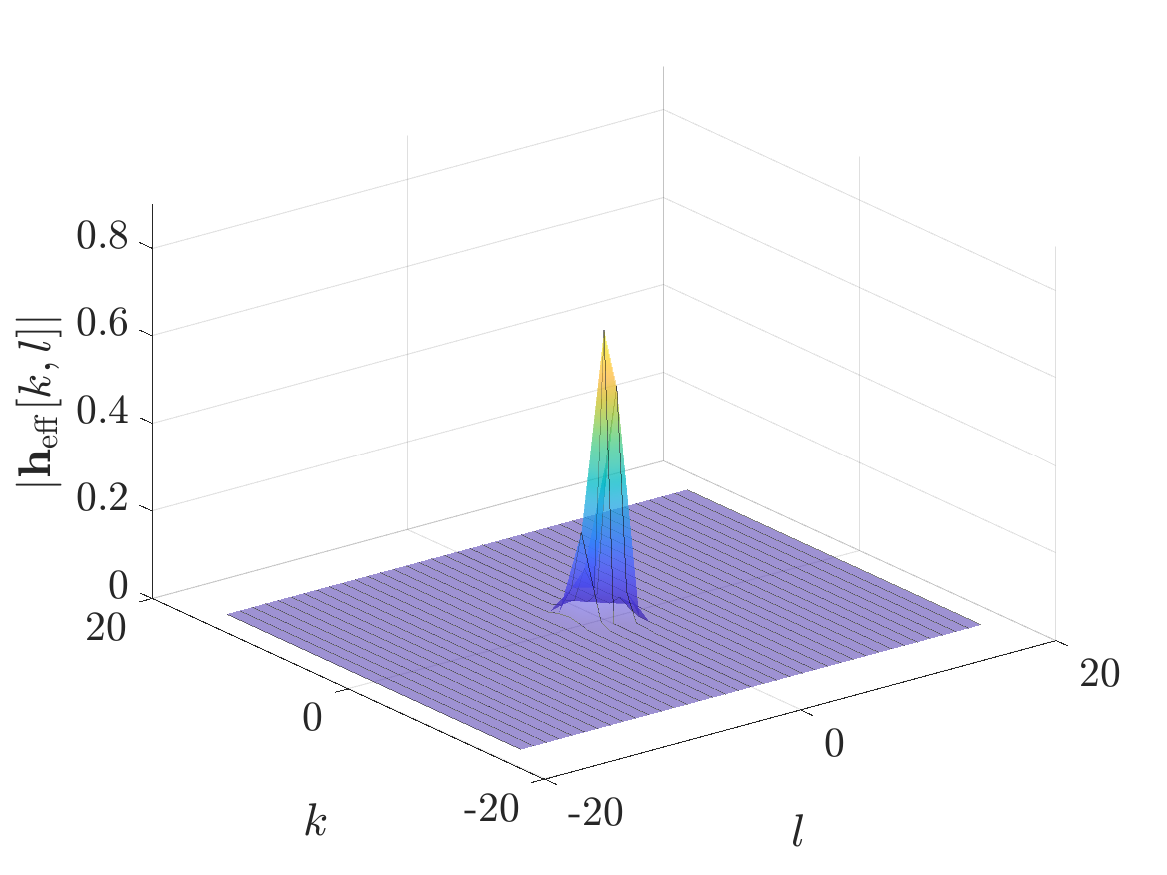}\label{fig:perfect_csi_31_37}}
    \hfill
    \subfloat[{Estimated $\mathbf{h}_{\mathrm{eff}}$ using DO with 4-QAM}]{\includegraphics[width=0.32\linewidth]{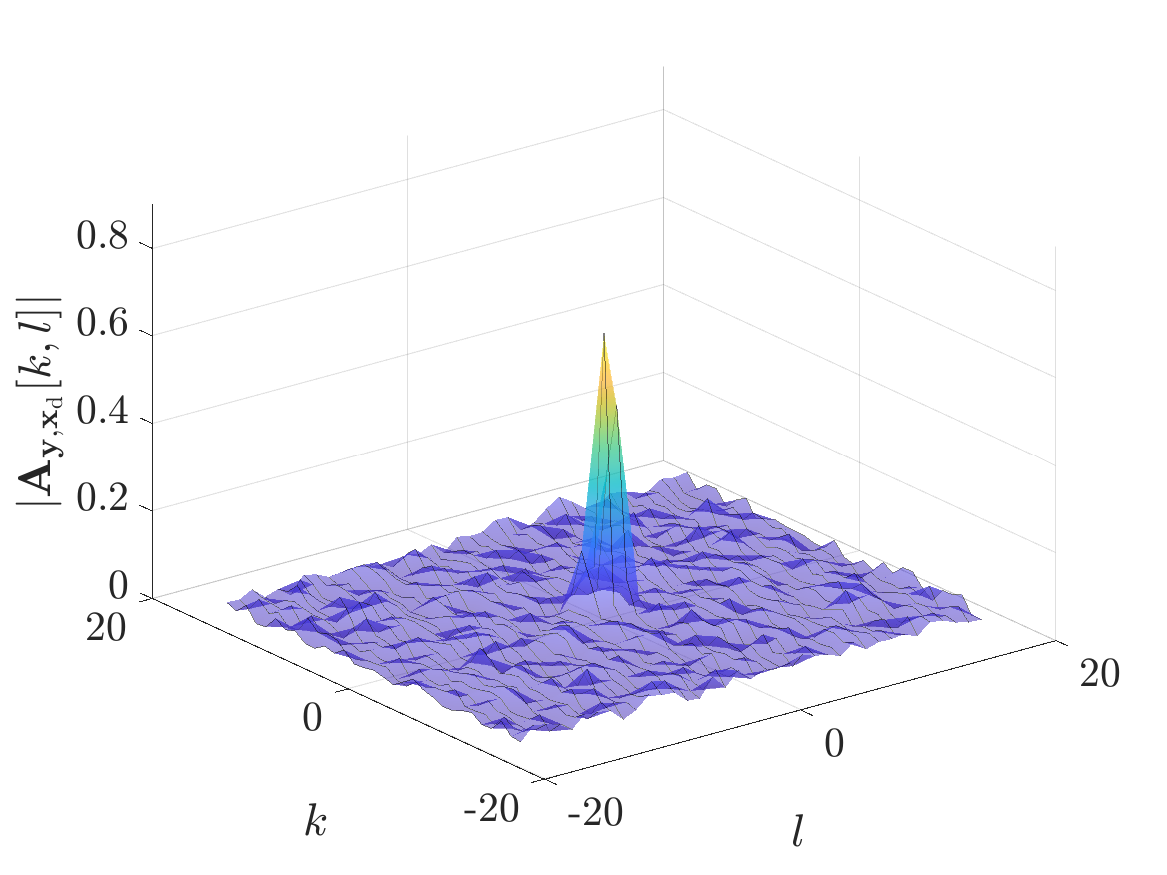}\label{fig:est_csi_31_37_4_qam}}
    \hfill
    \subfloat[{Estimated $\mathbf{h}_{\mathrm{eff}}$ using DO with 256-QAM}]{\includegraphics[width=0.32\linewidth]{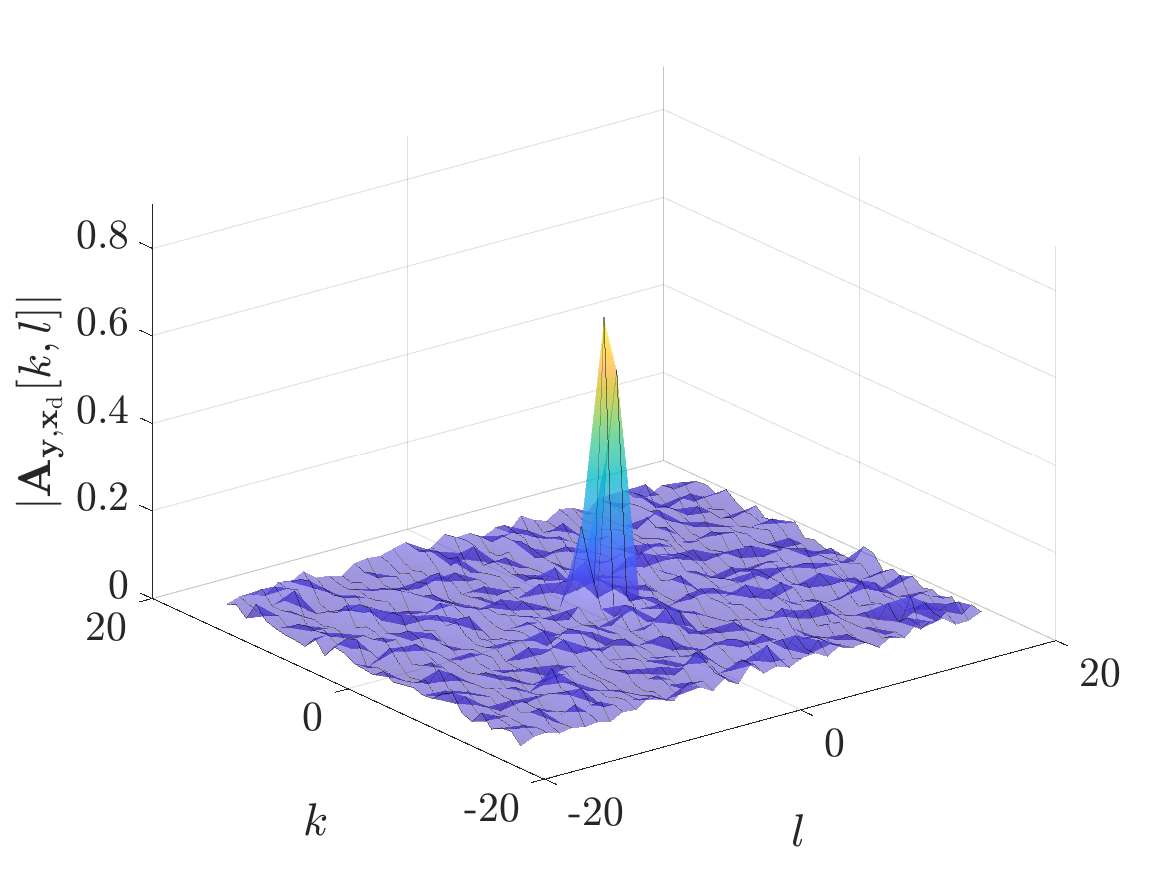}\label{fig:est_csi_31_37_256_qam}}\\
    \subfloat[{Actual $\mathbf{h}_{\mathrm{eff}}, M=101, N=103$}]{\includegraphics[width=0.32\linewidth]{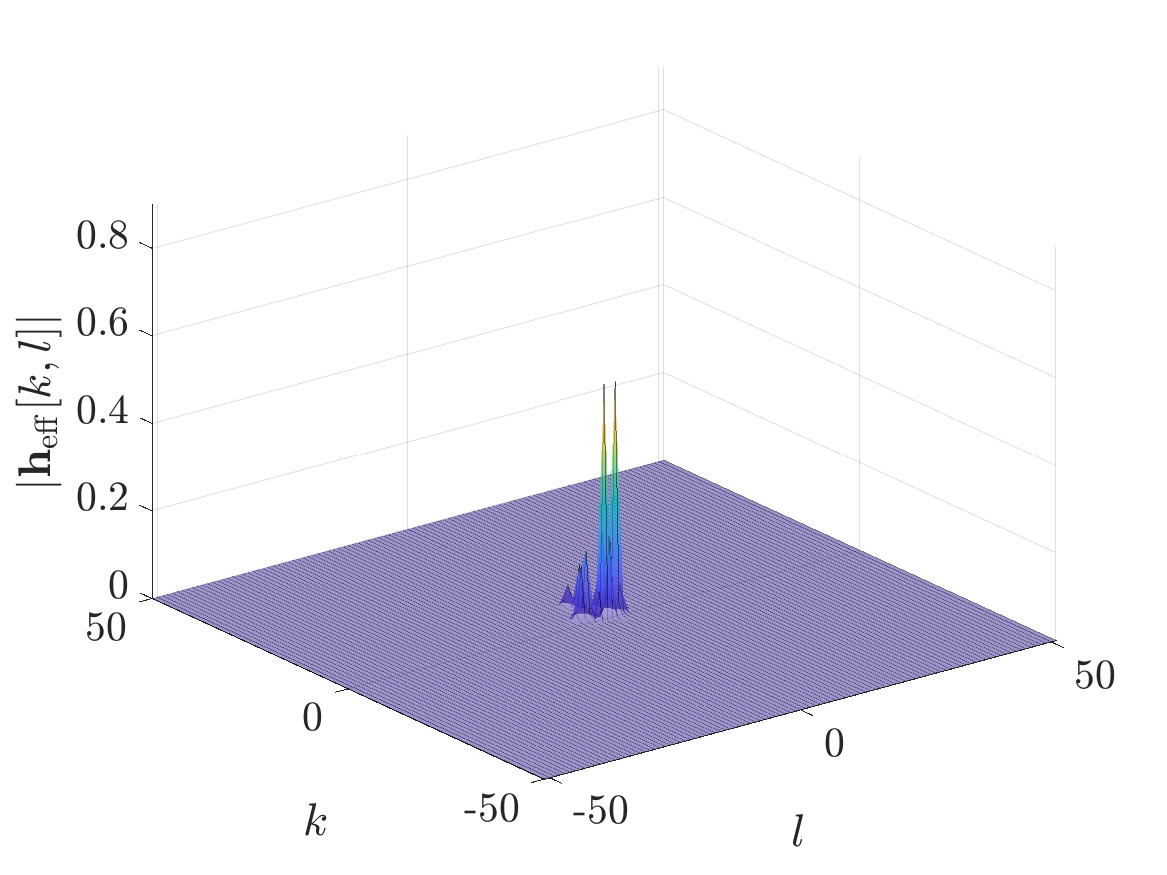}\label{fig:perfect_csi_101_103}}
    \hfill
    \subfloat[{Estimated $\mathbf{h}_{\mathrm{eff}}$ using DO with 4-QAM}]{\includegraphics[width=0.32\linewidth]{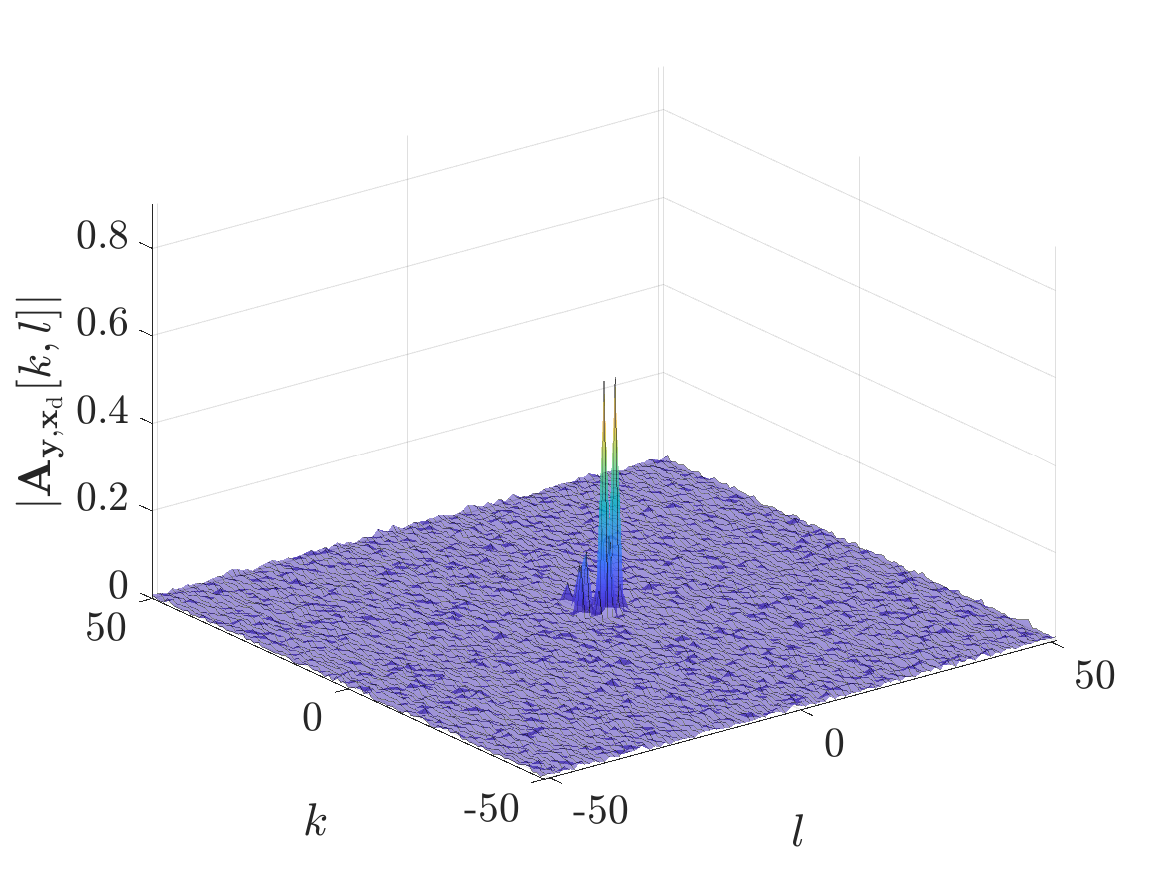}\label{fig:est_csi_101_103_4_qam}}
    \hfill
    \subfloat[{Estimated $\mathbf{h}_{\mathrm{eff}}$ using DO with 256-QAM}]{\includegraphics[width=0.32\linewidth]{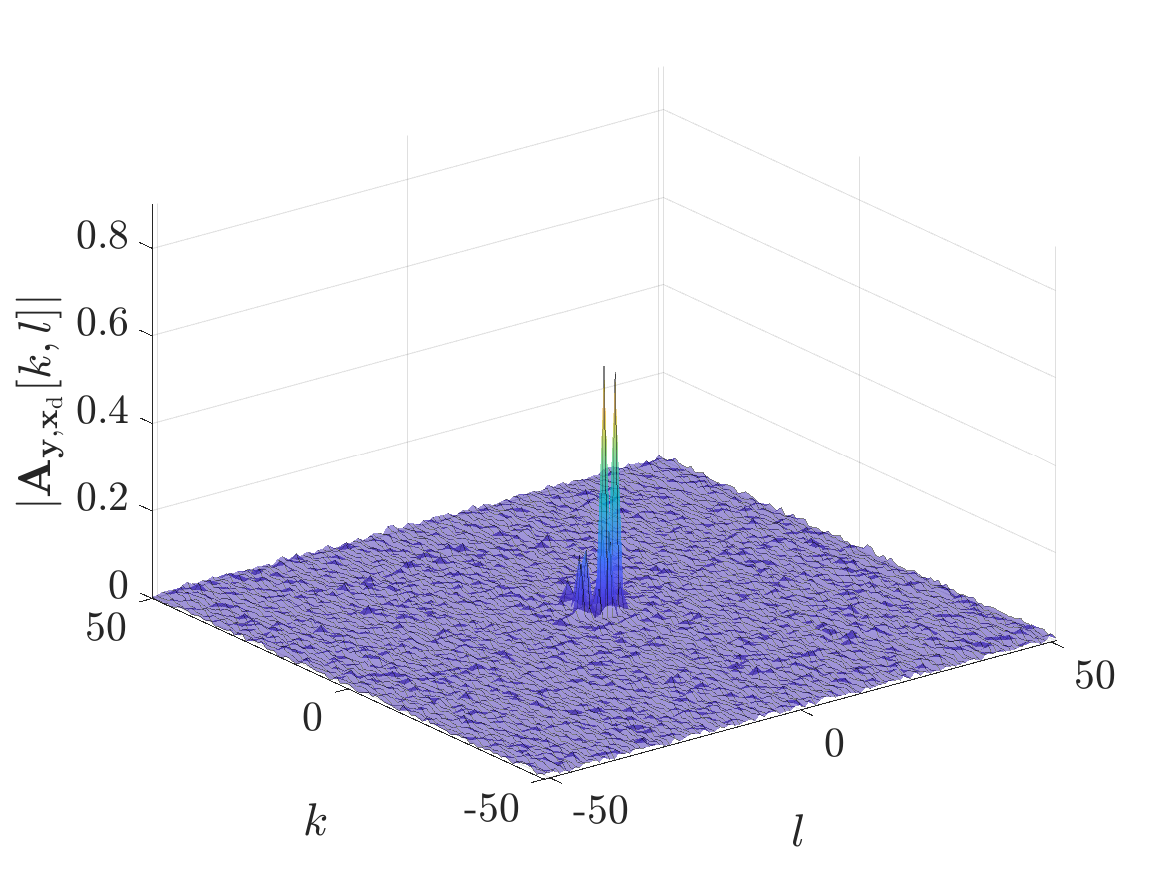}\label{fig:est_csi_101_103_256_qam}}
    \caption{Demonstrating the effect of size of frame and constellation on the performance of the proposed differential communication scheme for data SNR of 25 dB. RRC pulse shaping with parameter $\beta_\tau=\beta_\nu=0.6$. Three frame sizes are considered $M = 11, N = 13$ (first row), $M=31, N=37$ (second row), and $M=101, N=103$ (third row). For the DO frame, constellation sizes of 4-QAM and 256-QAM are considered. No significant performance difference between 4-QAM and 256-QAM across frame sizes. Channel estimation accuracy improves with increasing frame size.}
    \vspace{-4mm}
    \label{fig:effect_of_qam_frame_size}
\end{figure*}

\subsection{Error Propagation}

Figure \ref{fig:inst_nmse_no_of_frames} plots the instantaneous NMSE of the channel estimate obtained using the DO frame (using 4-QAM symbols) at the receiver at low and high SNR values. From the plot, it is observed that at low SNR (SNR = 0 dB) error propagation results in instantaneous NMSE increasing with the number of frames. It is also observed that every 30 frames, there is a dip in the NMSE value, wherein the estimate from the pilot frame curbs error propagation. However, at high SNR (SNR = 25 dB), the number of errors is low and there is no significant error propagation between two pilot transmissions. Note that, at SNR = 0 dB, the target NMSE required for performing close to perfect CSI is also very high, while at SNR = 25 dB, the target NMSE is lower to achieve close to perfect CSI performance and this trend is supported by the proposed receiver. 

\begin{figure*}
    \subfloat[{NMSE}]{\includegraphics[width=0.49\linewidth]{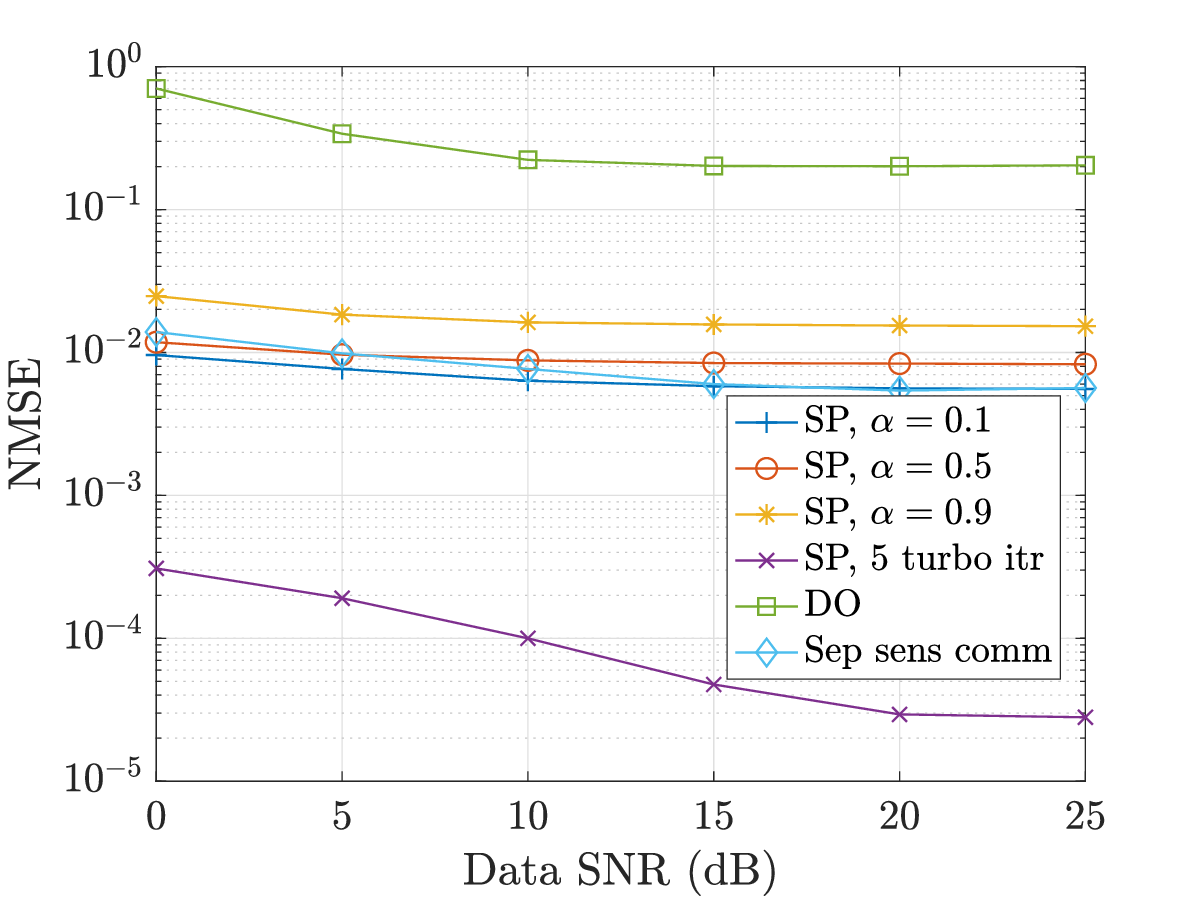}\label{fig:4_qam_nmse}}
    \hfill
    \subfloat[{BER}]{\includegraphics[width=0.49\linewidth]{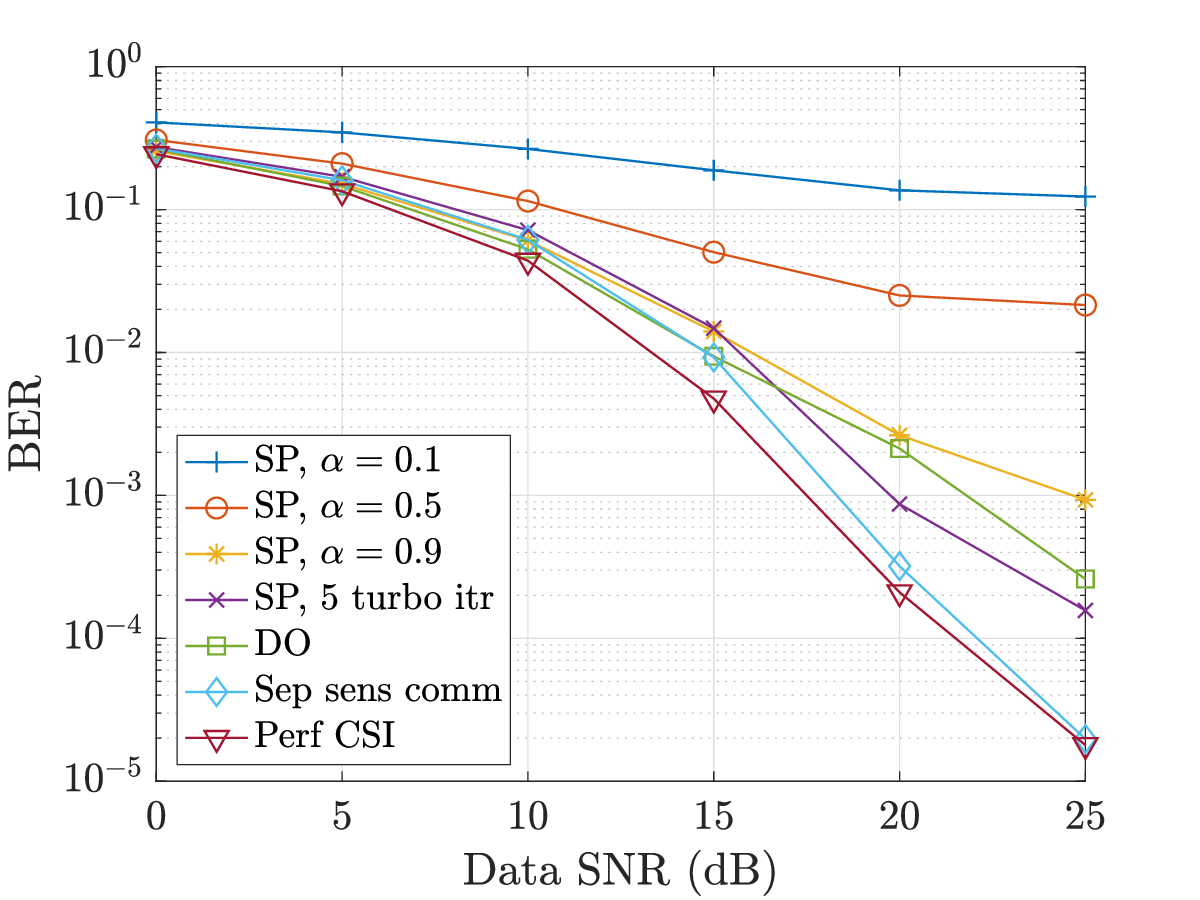}\label{fig:4_qam_ber}}
    \hfill
    \caption{Performance curves with proposed differential communication scheme, SP \cite{Aug2024paper} with varying power distribution, separate sensing and communication, and perfect CSI for 4-QAM modulation. Zak-OTFS system with $M=31, N=37, \nu_p=30 $ kHz, root raised cosine (RRC) pulse shaping with parameter $\beta_\tau=\beta_\nu=0.6$.}
    \vspace{-4mm}
    \label{fig:4_qam_performance}
\end{figure*}

\begin{figure*}
    \subfloat[{NMSE}]{\includegraphics[width=0.49\linewidth]{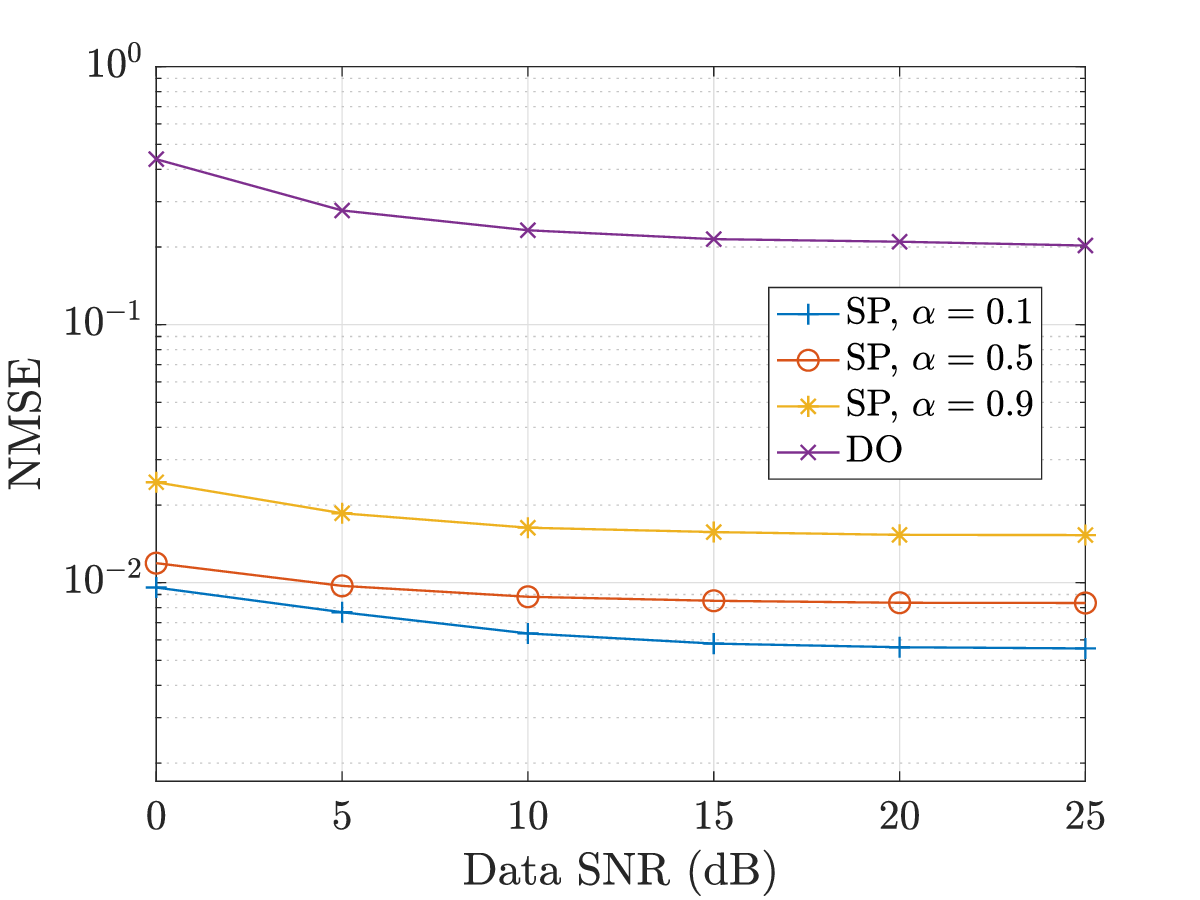}\label{fig:16_qam_nmse}}
    \hfill
    \subfloat[{BER}]{\includegraphics[width=0.49\linewidth]{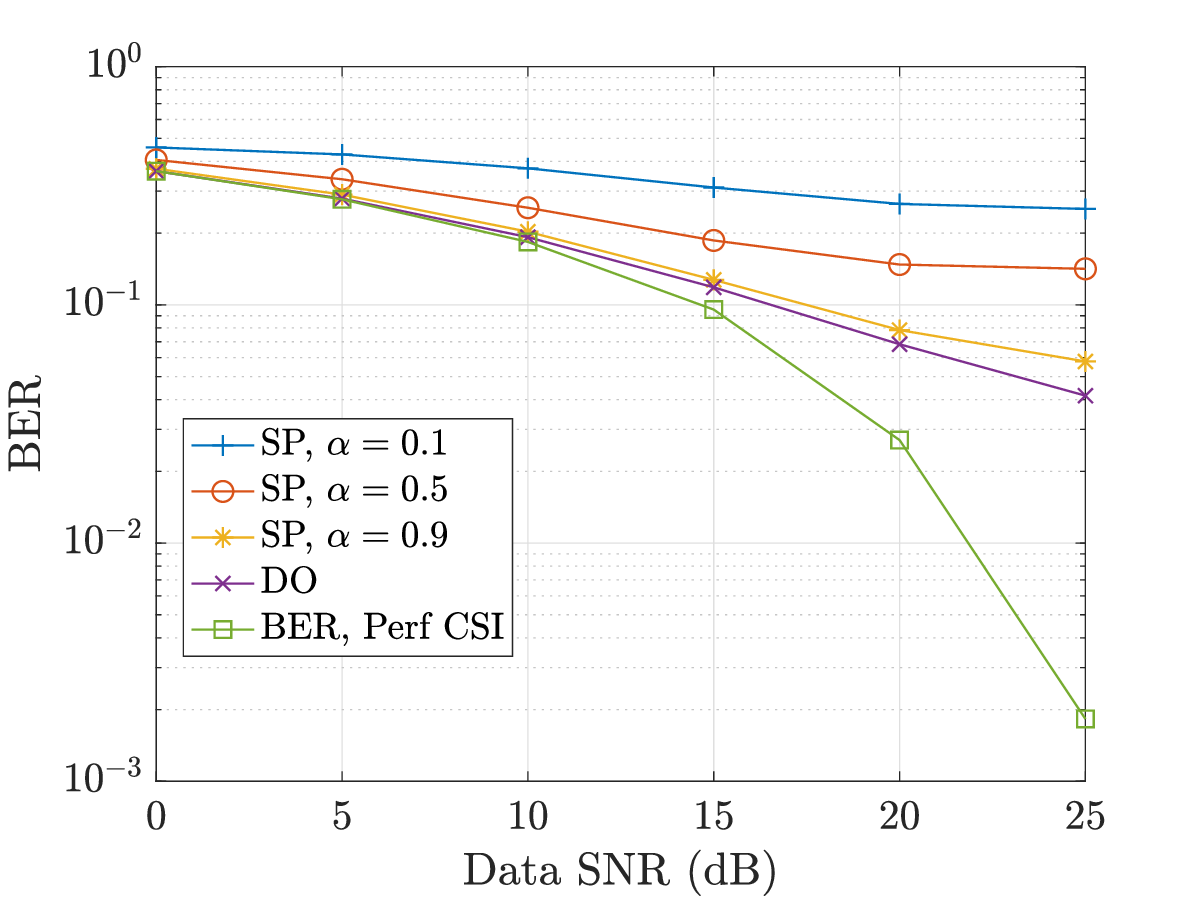}\label{fig:16_qam_ber}}
    \hfill
    \caption{Performance curves with proposed differential communication scheme, SP \cite{Aug2024paper} with varying power distribution, separate sensing and communication, and perfect CSI for 16-QAM modulation. Zak-OTFS system with $M=31, N=37, \nu_p=30 $ kHz, root raised cosine (RRC) pulse shaping with parameter $\beta_\tau=\beta_\nu=0.6$.}
    % \vspace{-4mm}
    \label{fig:16_qam_performance}
\end{figure*}

\subsection{Effect of Frame and Constellation Size on the Estimation Performance}
Figure \ref{fig:effect_of_qam_frame_size} plots the surface plot of the perfect channel and channel estimated using the DO frame for different frame sizes and constellation sizes. We consider a small frame with $M=11, N = 13$, a medium frame with $M=31, N = 37$, and a large frame with $M=101, N=103$. For each frame size, the channel estimated using the DO frame using 4-QAM (small constellation size) and 256-QAM (large constellation size) is plotted. It is seen that, for all frame sizes, the estimated channel does not vary significantly with change in constellation size but there is significant improvement with increase in frame size. This corroborates \eqref{eq:dif_det7}, wherein the estimate of the channel converges to the true channel asymptotically with frame size.

\subsection{4-QAM}
\label{subsec:4_qam}
Figure \ref{fig:4_qam_performance} shows the NMSE and BER performance of the Zak-OTFS system for various schemes using 4-QAM modulation. For the SP frame we consider various data and pilot distributions parameterized by $\alpha$. Keeping the total energy in the frame to be $e_\mathrm{do, lin}$, we consider $e_\mathrm{d, lin} = \alpha e_\mathrm{do, lin}$ and $e_\mathrm{p, lin} = (1-\alpha)e_\mathrm{do, lin}$ for various $\alpha$ values. The NMSE performance is seen to improve as $\alpha$ is decreased from $0.9$ to $0.1$. This is expected as the energy in the pilot increases leading to better estimates. SP frame with 5 turbo iterations \cite{jayachandran2024zakturbo} achieves the best NMSE as it involves cycling between channel estimation and data detection. The separate sensing and communication scheme involves transmitting separate pilot and data frames and the estimate from the pilot frame is used to detect data in the data frame. The NMSE of the proposed receiver with the DO frame floors at $0.2$. Moving to the BER performance, separate sensing and communication achieves performance closest to that with perfect channel state information (CSI), followed by the SP frame with turbo iterations. The performance of the proposed receiver using the DO frame is better than SP frames with various $\alpha$ values. It is interesting to note that the performance with DO frame is the lower bound for the performance with SP frame. \\
\textit{Complexity:} Here we compare the complexity qualitatively for full spectral efficiency achieving methods only. Separate sensing and communication does not achieve full spectral efficiency. The complexity is the highest for SP frame with turbo iterations owing to repeated channel estimation and data detection. For the SP frame without turbo iterations, there is an additional step involved to remove the contribution of pilot symbols before data is detected. The DO frame incurs the least complexity because there is no additional interference cancellation step. The DO frame therefore achieves better performance than SP at lower complexity.

\subsection{16-QAM}
\label{subsec:16_qam}
Figure \ref{fig:16_qam_performance} shows the NMSE and BER performance of the Zak-OTFS system with 16-QAM modulation. NMSE and BER trends are similar to that observed in 4-QAM. The SP frames with decreasing $\alpha$ values achieve better NMSE, while the DO frame achieves the best BER performance closest to the performance with perfect CSI, while being the least complex.

% # Compute pilot SNR
% snr_p = pdr + snr_d

% # Energies
% e_d = 10**(snr_d/10)
% e_p = 10**(snr_p/10)
% e_d_data_only = e_d + e_p

\section{Conclusions}
\label{sec:conclusions}
In this paper, we proposed a novel differential communication scheme for Zak-OTFS systems. The proposed scheme leveraged the predictability of the DD channel in Zak-OTFS to reduce the frequency of pilot symbol transmissions. The channel estimate obtained at a time instant was used for data detection in the next time instant. This allowed the information symbols transmitted in a DO frame to be used as pilots thereby achieving full spectral efficiency. We analytically showed the the cross-ambiguity between the received data frame and transmitted data frame provided a model-free estimate of the channel. Simulations results showed that the proposed detector using the DO frame achieved good NMSE and BER performance. SP and DO frames both achieve full spectral efficiency, but the proposed scheme with the DO frame achieved better BER performance at lower complexity than the SP frame.

\bibliographystyle{IEEEtran}
\bibliography{references}

% \clearpage
% \input{responses}
\end{document}